\documentclass[a4paper,fleqn,usenatbib]{mnras}

\usepackage[T1]{fontenc}
\usepackage{ae,aecompl}

\usepackage{graphicx}	
\usepackage{amsmath}	
\usepackage{amssymb}	

\title[]{Line-driven disk wind model for ultra-fast outflows in active galactic nuclei -- Scaling with luminosity}

\author[M. Nomura \& K. Ohsuga]{
M. Nomura,$^{1}$\thanks{E-mail: mariko.nomura@keio.jp}
and K. Ohsuga$^{2,3}$
\\
$^{1}$Department of Physics, Institute of Science and Technology, Keio University, 3-14-1 Hiyoshi, Yokohama, Kanagawa 223-8522, Japan\\
$^{2}$National Astronomical Observatory of Japan, 2-21-1 Osawa, Mitaka, Tokyo 181-8588, Japan\\
$^{3}$School of Physical Sciences, Graduate University of
Advanced Study (SOKENDAI), Shonan Village, Hayama, Kanagawa 240-0193, Japan
}

\date{Accepted XXX. Received YYY; in original form ZZZ}

\pubyear{2015}

\begin{document}
\label{firstpage}
\pagerange{\pageref{firstpage}--\pageref{lastpage}}
\maketitle

\begin{abstract}
In order to reveal the origin of the ultra-fast outflows (UFOs) 
that are frequently observed in active galactic nuclei (AGNs), 
we perform two-dimensional radiation hydrodynamics simulations 
of the line-driven disk winds, 
which are accelerated by the radiation force due to the spectral lines. 
The line-driven winds are successfully launched 
for the range of $M_{\rm BH}=10^{6-9}M_\odot$ and $\varepsilon=0.1$--$0.5$,
and the resulting mass outflow rate ($\dot{M_{\rm w}}$), 
momentum flux ($\dot{p_{\rm w}}$), 
and kinetic luminosity ($\dot{E_{\rm w}}$) 
are in the region containing 
90\% of the posterior probability distribution 
in the $\dot M_{\rm w}$-$L_{\rm bol}$ plane, 
$\dot p_{\rm w}$-$L_{\rm bol}$ plane,
and $\dot E_{\rm w}$-$L_{\rm bol}$ plane shown in Gofford et al.,
where 
$M_{\rm BH}$ is the black hole mass,  
$\varepsilon$ is the Eddington ratio, and
$L_{\rm bol}$ is the bolometric luminosity.
The best-fitting relations in Gofford et al.,
$d\log\dot{M_{\rm w}}/d\log{L_{\rm bol}}\sim 0.9$, 
$d\log\dot{p_{\rm w}}/d\log{L_{\rm bol}}\sim 1.2$, 
and $d\log\dot{E_{\rm w}}/d\log{L_{\rm bol}}\sim 1.5$,
are roughly consistent with our results, 
$d\log\dot{M_{\rm w}}/d\log{L_{\rm bol}}\sim 9/8$, 
$d\log\dot{p_{\rm w}}/d\log{L_{\rm bol}}\sim 10/8$, 
and $d\log\dot{E_{\rm w}}/d\log{L_{\rm bol}}\sim 11/8$.
In addition, 
our model predicts that no UFO features are detected 
for the AGNs with $\varepsilon \lesssim 0.01$,
since the winds do not appear.
Also, only AGNs with $M_{\rm BH} \lesssim 10^8 M_\odot$ exhibit the UFOs
when $\varepsilon \sim 0.025$.
These predictions nicely agree with the X-ray observations (Gofford et al.). 
These results support that the line-driven disk wind is
the origin of the UFOs.
\end{abstract}




\section{Introduction}
Blueshifted absorption lines of highly ionized iron
(FeX\hspace{-.1em}X\hspace{-.1em}V and/or 
FeX\hspace{-.1em}X\hspace{-.1em}V\hspace{-.1em}I) 
are observed
in some active galactic nuclei (AGNs).
The large blueshift of the iron lines
indicates the outflows blown away toward an observer
with velocity of $0.1$--$0.3c$,
where $c$ is the speed of light. 
Such high-velocity outflows,
which are so-called ultra-fast outflows (UFOs),
are detected in $\sim 40\%$ of Seyfert galaxies,
although the number of samples is not so large
\citep{Tombesi10,Tombesi11,Gofford13}.

The UFOs are thought to be the disk winds, 
which are launched from the accretion disks
around supermassive black holes, 
but the acceleration mechanism is still unknown. 
One plausible model is the line-driven wind model \citep{Proga00,Proga04,Risaliti10,Nomura13},
in which the acceleration mechanism is radiation force due to spectral lines 
(the absorbing the ultraviolet radiation 
through the bound-bound transition of metals).
In this model, 
the line force becomes 10--1000 times larger than the radiation force 
due to Thomson scattering near the disk surface
where the ionization state is low \citep{SK90},
leading to the launch of the high-velocity disk winds.
At the region far from the disk surface,
since the wind matter is highly ionized by the X-ray irradiation
from the disk corona,
the blueshifted absorption lines of highly ionized iron lines are
produced.
The magnetically driven wind model
\citep[e.g,][]{B82,K94,Ever07,Fukumura15},
which is another theoretical model,
can explain the large velocity of the outflowing matter,
but this model requires an extra mechanism 
to adjust the ionization state of the wind matter
to explain the absorption lines.

Radiation hydrodynamics simulations of the line-driven wind in AGNs 
have been developed by \citet{Proga00} and \citet{Proga04}.
They clearly show that the line force accelerates the matter and 
the funnel-shaped disk wind appears for the typical parameters, 
$M_{\rm BH}=10^8 M_{\odot}$ and $\varepsilon=0.5$, 
where $M_{\rm BH}$ and $\varepsilon$ are the black hole mass
and Eddington ratio of the disk luminosity respectively.
In their simulations,
the matter is launched from the disk surface
and is accelerated toward the direction of $\theta\sim 70^{\circ}$, 
where $\theta$ is the polar angle 
measured from the rotational axis of the accretion disk. 
Based on their results, \citet{Schurch09} and \citet{Sim10} performed 
the spectral synthesis 
and revealed that the blueshifted absorption lines appear on the spectra 
transmitted through the line-driven disk winds.

In addition, in our previous work \citep[][hereafter N16]{MN2016},
we also performed the radiation hydrodynamics simulations 
of line-driven disk winds 
in a wide parameter space and explained the observational features of the UFOs. 
For $M_{\rm BH} = 10^{6-9} M_\odot$ and $\varepsilon = 0.1$--$0.7$, 
funnel-shaped disk winds appear, 
and dense matter is accelerated outward with an outflow velocity
of $\sim 0.1c$ and with an opening angle of $70$--$80^{\circ}$. 
The outward velocity, the column density, and 
the ionization state of the winds
are consistent with those 
inferred from the X-ray observation of the UFOs. 
The UFOs could be statistically observed in about 13--28\%
of the luminous AGNs ($\varepsilon \gtrsim 0.1$), 
which is roughly comparable to the observed ratio ($\sim 40\%$). 

Recently, \citet{Gofford15} 
(hereafter G15)
surveyed the wind properties for 20 samples of UFOs in detail.
They found that the mass outflow rate ($\dot{M_{\rm w}}$), 
momentum flux ($\dot{p_{\rm w}}$), 
and kinetic luminosity ($\dot{E_{\rm w}}$) 
are scaling with AGN bolometric luminosity ($L_{\rm bol}$) as
$\log\dot{M_{\rm w}}\sim 0.9 \log{L_{\rm bol}}-13$, $\log\dot{p_{\rm w}}\sim 1.2 \log{L_{\rm bol}} -18.1$, and $\log\dot{E_{\rm w}}\sim 1.5 \log{L_{\rm bol}}-23.5$.
The line-driven disk wind model should explain this scaling relation 
if this wind is the origin of the UFOs.
However, the above scaling of the line-driven disk wind
has not been investigated yet.

In this paper, by performing two-dimensional
radiation hydrodynamics simulations
of the line-driven disk winds,
we research the dependence of the mass outflow rate, momentum flux, 
and kinetic luminosity on the disk luminosity
(that corresponds to the bolometric luminosity).
In Section \ref{method}, we explain the calculation method,
which is basically the same as N16 except for the slight modification.
Our results are shown in Section \ref{result}.
We devote Section \ref{discussions} to discussions and we present conclusions in Section \ref{conclusions}. 



\section{Methods}
\label{method}

In this section,
we briefly describe our method, which is basically the same as that of N16, 
and mention changes from the previous simulations.
We apply the spherical polar coordinate $(r,\,\theta,\,\phi)$, 
where $r$ is the distance from the origin of the coordinate, 
$\theta$ is the polar angle, 
and $\phi$ is the azimuthal angle. 
We perform the two-dimensional simulations 
assuming the axial symmetry for the rotation axis of the accretion disk.
We calculate the basic equations of the hydrodynamics,
the equation of continuity,
\begin{equation}
\frac{\partial \rho}{\partial t}+\nabla\cdot(\rho \mbox{\boldmath $v$})
=0,
\label{eoc}
\end{equation}
the equations of motion,
\begin{equation}
\frac{\partial (\rho v_r)}{\partial t}+\nabla\cdot(\rho v_r \mbox{\boldmath $v$})
=-\frac{\partial p}{\partial r}+\rho\Bigg[\frac{v_\theta^2}{r}+\frac{v_\varphi^2}{r}+g_r+f_{{\rm rad},\,r}\Bigg],
\label{eom1}
\end{equation}
\begin{equation}
\frac{\partial (\rho v_\theta)}{\partial t}+\nabla\cdot(\rho v_\theta \mbox{\boldmath $v$})
=-\frac{1}{r}\frac{\partial p}{\partial \theta}+\rho\Bigg[-\frac{v_r v_\theta}{r}+\frac{v_\varphi^2}{r}\cot \theta+g_\theta+f_{{\rm rad},\,\theta}\Bigg],
\label{eom2}
\end{equation}
\begin{equation}
\frac{\partial (\rho v_\varphi)}{\partial t}+\nabla\cdot(\rho v_\varphi \mbox{\boldmath $v$})
=-\rho\Bigg[\frac{v_\varphi v_r}{r}+\frac{v_\varphi v_\theta}{r}\cot \theta\Bigg],
\label{eom3}
\end{equation}
and the energy equation,
\begin{equation}
\frac{\partial}{\partial t}\Bigg[  \rho \Bigg(\frac{1}{2}v^2+e\Bigg) \Bigg]
+\nabla \cdot \Bigg[  \rho\mbox{\boldmath $v$} \Bigg(\frac{1}{2}v^2+e+\frac{p}{\rho}\Bigg) \Bigg]
=\rho\mbox{\boldmath $v$}\cdot\mbox{\boldmath $g$}+\rho\mathcal L,
\label{eoe}
\end{equation}
where $\rho$ is the mass density, 
\mbox{\boldmath $v$}$=(v_r,\,v_\theta,\,v_\varphi)$ is the velocity, $p$ is the gas pressure,
$e$ is the internal energy per unit mass, 
and \mbox{\boldmath $g$}$=(g_r,\,g_\theta )$ is the gravitational acceleration of the black hole.
The $\theta$-component of the gravitational force is not null 
because the center of the coordinate and the $\theta=\pi/2$ plane are located 
at a distance of $z_0$ above the black hole and 
the mid-plane of the accretion disk.
We assume an adiabatic equation of state $p/\rho=(\gamma -1)e$ 
with $\gamma=5/3$. 
In the last term of the equation (\ref{eoe}), 
$\mathcal L$ is the net cooling rate.
Here we consider Compton heating/cooling, 
X-ray photoionization heating, recombination cooling, 
bremsstrahlung cooling, and line cooling (see N16).

In equations (\ref{eom1}) and (\ref{eom2}), the last term,
\mbox{\boldmath $f$}$_{\rm rad}=(f_{{\rm rad},\,r},\,f_{{\rm rad},\,\theta})$, 
is the radiation force per unit mass including the line force, 
which is calculated as 
\begin{equation}
{ \mbox{\boldmath $f$}_{\rm rad}}=\frac{\sigma_{\rm e} \mbox{\boldmath $F$}_{\rm D}}{c}+\frac{\sigma_{\rm e} \mbox{\boldmath $F$}_{\rm line}}{c}M,
 \label{radforce}
\end{equation}
where $\sigma_{\rm e}$ is the mass-scattering coefficient for free electrons, 
\mbox{\boldmath $F$}$_{\rm D}$ is the radiation flux via the disk emission, 
and $M$ is the force multiplier proposed by \citet{SK90}.
In addition,
\mbox{\boldmath $F$}$_{\rm line}$ is the radiation flux of the disk emission
in a band of $200$--$3200\,{\rm \AA}$,
which largely contributes to the line force
through the bound-bound transitions
\citep[line-driving flux, see][]{Proga04}.
In N16,
we roughly assume that the radiation from the high-temperature region
of the disk,
in which the effective temperature is larger than $3\times 10^3\,{\rm K}$,
contributes to the line force \citep[see also][]{Proga00}.
The line force would be estimated more precisely
by the present method.

We also modify the evaluation method of the velocity gradient 
when we estimate the force multiplier, $M$.
The force multiplier defined by \citet{SK90}
is the function of the ionization parameter,
\begin{equation} 
 \xi=\frac{4\pi F_{\rm X}}{n},
  \label{xi}
\end{equation}
and the local optical depth parameter,
\begin{equation}
  t=\sigma_{\rm e} \rho v_{\rm{th}}\Bigl| \frac{dv}{ds}\Bigr|^{-1},
   \label{t-xi}
\end{equation} 
where $F_{\rm X}$ is the X-ray flux
from the disk corona near the black hole,
$n$ is the number density,
$v_{\rm{th}}$ is the thermal speed of hydrogen gas
whose temperature is $25,000\,{\rm K}$ ($v_{\rm th}=20\,{\rm km\,s^{-1}}$),
and $dv/ds$ is the velocity gradient along the light-ray.
In this paper, we evaluate $dv/ds$ by the velocity gradient
along the direction of the line-driving flux
(\mbox{\boldmath $F$}$_{\rm line}/|$\mbox{\boldmath $F$}$_{\rm line}|$),
although 
we substituted $dv_r/dr$ for $dv/ds$ in N16.
The present method would be better than the previous method,
since the light-ray along the direction of the line-driving flux
effectively contributes the line force.
For instance, around the wind base, 
the matter is accelerated almost perpendicularly to the disk plane
via the $\theta$-component of the radiation force.
Then, the present method ($dv/ds \sim dv_\theta/r d\theta$)
is more suitable than the previous method ($dv/ds = dv_r/dr$).
In the same way as for N16,
the X-ray source is treated as a point source
located at the origin.
Also, the X-ray flux, the disk flux, and
the line-driving flux are attenuated in the same manner as N16.

The size of the computational domain 
and the grid spacing are the same as those of N16.
However, 
the distance from the equatorial plane to the $\theta=\pi/2$ plane,
$z_0$, is slightly modified.
We set $z_0$ to be $3.1\varepsilon R_{\rm S}$,
which is the scale height of the standard disk model at $r=30R_{\rm S}$ \citep{SS73},
although $z_0= 4 \varepsilon R_{\rm S}$ was employed in N16.

The initial and boundary conditions are the same as those of N16,
except for the density distribution at $\theta=\pi/2$ plane,
$\rho(\theta=\pi/2)$.
For $\rho(\theta=\pi/2)$,
we employ the density at the surface of the standard accretion disk, 
\begin{equation}
\begin{array}{ll}
& \rho(\theta=\pi/2) =\bar \rho(M_{\rm BH},\varepsilon,r)/(e-1)\\
& =
\left\{
    \begin{array}{l}
     5.24\times10^{-4}(M_{\rm BH}/M_{\odot})^{-1}(\varepsilon/\eta)^{-2}(r/R_{\rm S})^{3/2}\,{\rm g\,cm^{-3}}\\
     \qquad \qquad \qquad \qquad r\leq 18(M_{\rm BH}/M_{\odot})^{2/21}(\varepsilon/\eta)^{16/21}R_{\rm S} \\
     4.66(M_{\rm BH}/M_{\odot})^{-7/10}(\varepsilon/\eta)^{2/5}(r/R_{\rm S})^{-33/20}\,{\rm g\,cm^{-3}} \\
     \qquad \qquad \qquad \qquad r > 18(M_{\rm BH}/M_{\odot})^{2/21}(\varepsilon/\eta)^{16/21}R_{\rm S}\\
    \end{array} \right.
    \end{array}
 ,
\end{equation}
where $\bar \rho(M_{\rm BH},\varepsilon,r)$ is the vertically averaged density 
of the standard disk model \citep{SS73,Kato_textbook} 
and $\eta$ is the energy conversion rate that is set to be $\eta=0.06$,
although we assumed $\rho(\theta=\pi/2)$ to be constant in N16.
Although we employ the above modifications,
the wind structure does not change so much.
The density of the wind slightly increases 
and the opening angle slightly decreases (see Section \ref{result1} for details).

\section{Results}
\label{result}
\subsection{Overview of the wind structure}
\label{result1}
\begin{figure}
  \begin{center}
    \includegraphics[width=\columnwidth]{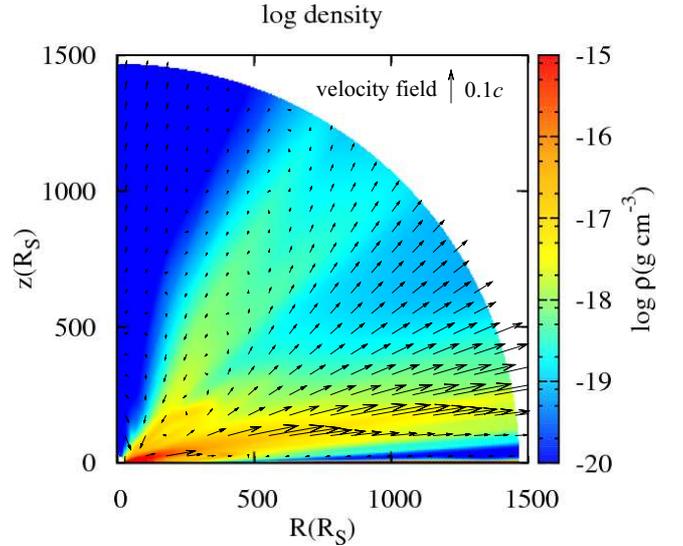}
  \end{center}
  \caption{
Time-averaged color density map of the line-driven disk wind 
for $M_{\rm BH}=10^8 M_{\odot}$ and $\varepsilon=0.1$. 
The vectors show the velocity field. 
The $z=0$ plane corresponds to the accretion disk surface 
and the $z$-axis is the rotational axis of the disk.
}\label{fig1}
\end{figure}

Here, we show the results of a model with 
$M_{\rm BH}=10^8 M_{\odot}$ and $\varepsilon=0.1$.
We call this model a ``fiducial model'' 
because a large sample of UFOs is detected in AGNs with $\varepsilon\sim 0.1$ in G15.
Fig. \ref{fig1} shows the time-averaged wind structure 
for the fiducial model in the $R$-$z$ plane. 
Here, $z$-axis is the rotation axis of the accretion disk and 
$R$ is the distance from the rotation axis.
The color contour shows the density map and arrows show the velocity vectors. 
The matter is launched from the accretion disk surface near the black hole 
($R=30$--$40R_{\rm S}$) almost 
vertically.
After the launching, 
the direction of acceleration bends in the $r$-direction and 
the outflowing matter makes the funnel-shaped disk wind
with an opening angle of $70$--$80^{\circ}$.
These results are consistent with N16.

Here we note that the disk wind is slightly 
denser and faster than that of N16.
The maximum velocity is 1.5--2.0 times larger 
and the density is 10--30 times larger 
than those of previous results. 
Such changes are mainly caused by the modification 
of the treatment of the velocity gradient
(see Section \ref{method}). 
In the launching region, 
the matter is vertically accelerated  
by the $z$-component ($\theta$-component) of the radiation force.
Therefore, the velocity gradient along the direction of the radiation flux 
is larger than that along the radial direction.
Since the line force (force multiplier)
increases as an increase in the velocity gradient,
the velocity as well as the density of the wind increases
in the present simulations.

\begin{figure}
  \begin{center}
    \includegraphics[width=\columnwidth]{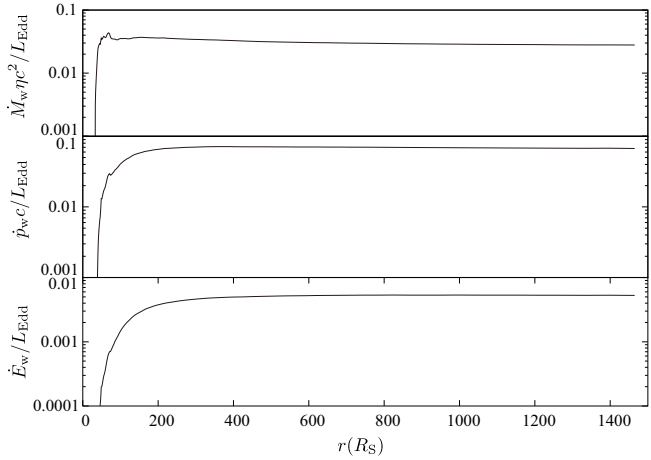}
  \end{center}
  \caption{
Mass outflow rate (top panel), 
momentum flux (middle panel), 
and kinetic luminosity (bottom panel) as a function of the distance from the center of the coordinate. 
All profiles are flat in the region of $r \gtrsim 100R_{\rm S}$ because the acceleration of the wind 
is almost terminated at $\sim 100R_{\rm S}$.
}\label{fig2}
\end{figure}

In Fig. \ref{fig2}, the mass outflow rate ($\dot M_{\rm w}$, top panel), 
momentum flux ($\dot p_{\rm w}$, middle panel), 
and kinetic luminosity ($\dot E_{\rm w}$, bottom panel)
are plotted as a function of $r$.
Here they are evaluated as
\begin{equation}
 \dot M_{\rm w}=4\pi r^2 \int^{89^{\circ}}_0 
  \rho v_r \sin\theta d \theta,
\end{equation}
\begin{equation}
 \dot p_{\rm w}=4\pi r^2 \int^{89^{\circ}}_0 
  \rho v_r^2 \sin\theta d \theta,
\end{equation}
and
\begin{equation}
 \dot E_{\rm w}=4\pi r^2 \int^{89^{\circ}}_0 
  \frac{1}{2}\rho v_r^3 \sin\theta d \theta,
\end{equation}
at the distance $r$. 
In order to avoid the influence of the boundary condition at $\theta=90^\circ$,
we set the $\theta$ range for the integration to be $0$--$89^\circ$.
The high-density and high-velocity part of the wind,
$\theta \sim 70$--$80^{\circ}$ (see Fig. \ref{fig1}),
is responsible for $\dot M_{\rm w}$, $\dot p_{\rm w}$, and $\dot E_{\rm w}$.

We find that all the profiles are steep in the region of 
$r\lesssim 100R_{\rm S}$
and nearly constant 
in the region of $r \gtrsim 100R_{\rm S}$,
since the wind is efficiently accelerated within $r\sim 100R_{\rm S}$
and since the acceleration of the wind 
is almost terminated at $\sim 100R_{\rm S}$.
Hereafter, 
we use the values at the outer boundary ($r=1500 R_{\rm S}$)
as the mass outflow rate, momentum flux, and the kinetic luminosity 
of the disk wind.
Fig. \ref{fig2} indicates 
$\dot{M}_{\rm w}\sim 5.0\times 10^{24}\,{\rm g\,s^{-1}}$, 
$\dot{p}_{\rm w}\sim 2.0\times 10^{34}\,{\rm g\, cm\, s^{-2}}$, 
and $\dot{E}_{\rm w}\sim 4.7 \times 10^{43}\,{\rm erg\, s^{-1}}$
in the fiducial model.
We find that the mass outflow rate is $\sim 22\%$ 
of the mass accretion rate, 
$\dot{M}_{\rm acc}=\varepsilon L_{\rm Edd}/\eta c^2$, 
where 
$L_{\rm Edd}$ is the Eddington luminosity. 
The kinetic luminosity is $\sim 3.8\%$ of the disk luminosity ($L_{\rm D}=\varepsilon L_{\rm Edd}$).

\subsection{Scaling with luminosity}
\begin{figure}
  \begin{center}
    \includegraphics[width=\columnwidth]{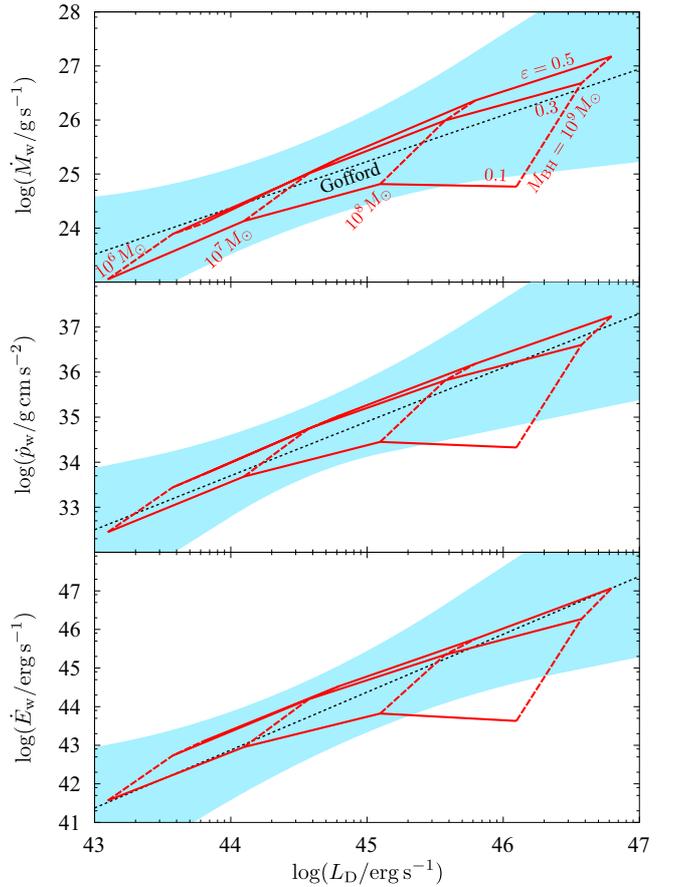}
  \end{center}
  \caption{Mass outflow rate (top panel), 
momentum flux (middle panel), 
and kinetic luminosity (bottom panel) 
as a function of the disk luminosity.
In each panel, the red solid lines show the results for fixed Eddington ratios,
$\varepsilon=0.1$, 0.3, and 0.5. 
The red dashed lines show the results for fixed black hole masses,
$M_{\rm BH}=10^6 M_{\odot}$, $10^7 M_{\odot}$, $10^8 M_{\odot}$, and $10^9 M_{\odot}$.
The 
black dotted 
line and the blue shaded area correspond to
the best-fitting regression line and the area containing 90\% of the posterior probability distribution,
which is evaluated by the observational data in G15.
}
\label{fig3}
\end{figure}

Fig. \ref{fig3} shows the mass outflow rate (top panel), 
momentum flux (middle panel), 
and kinetic luminosity (bottom panel) 
as a function of the disk luminosity. 
In each panel, 
the lines for fixed Eddington ratios, $\varepsilon=0.1$, 0.3, and 0.5, 
are shown by red solid lines.
The lines for fixed black hole masses, $M_{\rm BH}=10^6 M_{\odot}$, $10^7 M_{\odot}$, $10^8 M_{\odot}$, and $10^9 M_{\odot}$, are shown by red dashed lines. 
Here, we note that 
the range of $M_{\rm BH}$ and $\varepsilon$ employed in the present work
($M_{\rm BH}=10^{6-9}M_\odot$ and $\varepsilon=0.1$--$0.5$)
is roughly consistent with that of the sample of G15. 
The 
black dotted
lines are functions of $\log\dot{M_{\rm w}}\sim 0.9 \log{L_{\rm bol}}-13$ (top),
$\log\dot{p_{\rm w}}\sim 1.2 \log{L_{\rm bol}} -18.1$ (middle), 
and $\log\dot{E_{\rm w}}\sim 1.5 \log{L_{\rm bol}}-23.5$ (bottom).
The upper and lower boundary curves of the blue shaded area are 
the envelopes of the lines, 
$\log \dot{M_{\rm w}}=a \log L_{\rm bol}-1.03a^2-42.4a+24.5$ 
and $\log \dot{M_{\rm w}}=a \log L_{\rm bol}+1.20a^2-46.7a+25.0$
with $a$ being distributed in the range of $0.286\le a \le 1.60$
(top panel),
$\log \dot{p_{\rm w}}=a \log L_{\rm bol}-0.931a^2-42.2a+33.7$ 
and $\log \dot{p_{\rm w}}=a \log L_{\rm bol}+0.448a^2-45.7a+34.4$ 
with $a$ being distributed in the range of $0.625\le a \le 1.91$
(middle panel), 
and $\log \dot{E_{\rm w}}=a \log L_{\rm bol}-0.754a^2-42.1a+42.7$ 
and $\log \dot{E_{\rm w}}=a \log L_{\rm bol}+0.901a^2-47.7a+45.3$
where $0.714\le a \le 2.35$
(bottom panel).
Using the 
black dotted
lines and shaded areas, we can 
trace the best-fitting regression line 
and the region containing 90\% of the posterior probability distribution,
which are shown by the black lines and by the grey shaded area 
in Fig. 3 of G15.
Here, we assume that the luminosity of the accretion disk corresponds 
to the bolometric luminosity of the AGN.
In each panel, we find that our results nicely fit the shaded region,
although only the case of $\varepsilon=0.1$ and $M_{\rm BH}=10^9M_\odot$
is slightly protruded.
Also, 
our results are almost overlapped with the best-fitting lines.
This implies that 
the line-driven wind model can explain 
the observations of UFOs.

As shown in Fig. \ref{fig3},
the gradient of lines for fixed Eddington ratio 
($\varepsilon=0.3$ and $0.5$) is 
very close to that of best-fitting line,
while the red dashed lines are steeper than 
the black dotted lines.
This means that the agreement between our model
and the observations by G15
is mainly caused by the $M_{\rm BH}$-dependency of 
$\dot{M_{\rm w}}$, $\dot{p_{\rm w}}$, and $\dot{E_{\rm w}}$.
Note that 
the gradient of the line of $\varepsilon=0.1$ 
is mostly the same as that of $\varepsilon=0.3$ and $0.5$
($M_{\rm BH} \leq 10^8M_\odot$),
but much flatter when $M_{\rm BH}>10^8M_\odot$.
We will discuss that (see below).

If we plot the results of $0.025 \leq \varepsilon < 0.1$
and $0.5 < \varepsilon \leq 0.7$, 
the area in which our results are distributed slightly expands downward.
The results for $0.5 < \varepsilon \leq 0.7$ 
almost overlap with 
the red solid line of $\varepsilon =0.5$.
When $\varepsilon=0.025$,
the results of $\dot{M}_{\rm w}$, $\dot{p}_{\rm w}$, and $\dot{E}_{\rm w}$ 
for $M_{\rm BH} \leq 10^7 M_\odot$ are comparable to or slightly smaller than the lower boundary of 
the blue shaded area, and the disk wind does not appear for $M_{\rm BH} \geq 10^8 M_\odot$.
However, the scaling relation
does not change so much.
In addition, the sample of G15 is 
clustered at $\varepsilon=0.1$. 
Thus, we stress again that 
our model can nicely reproduce 
the observed scaling relation of 
$\dot{M}_{\rm w}$, $\dot{p}_{\rm w}$, and $\dot{E}_{\rm w}$ 
with the luminosity.

\subsection{Reason for scaling}
As we have mentioned above,
the scaling relation with the bolometric luminosity
is principally responsible for the black hole mass dependence.
Thus, in this subsection,
we discuss the black hole mass dependence of 
$\dot{M}_{\rm w}$, $\dot{p}_{\rm w}$, and $\dot{E}_{\rm w}$.

\begin{figure}
  \begin{center}
    \includegraphics[width=\columnwidth]{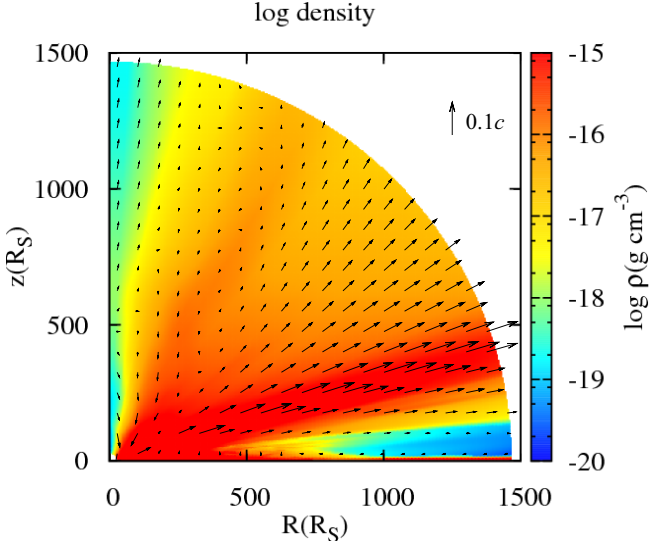}
  \end{center}
  \caption{
Same as Fig. \ref{fig1}, but for $M_{\rm BH}=10^6 M_{\odot}$.
}\label{fig4}
\end{figure}

Fig. \ref{fig4} is the same as Fig. \ref{fig1} 
but for $M_{\rm BH}=10^6 M_{\odot}$. 
Compared to the wind for $M_{\rm BH}=10^8 M_{\odot}$, 
the density of the wind is large by two orders of magnitude 
and the outward velocity is slightly small.
In addition, the opening angle of the wind is 
slightly smaller for $M_{\rm BH}=10^6 M_{\odot}$
than for $M_{\rm BH}=10^8 M_{\odot}$. 

\begin{figure}
  \begin{center}
    \includegraphics[width=\columnwidth]{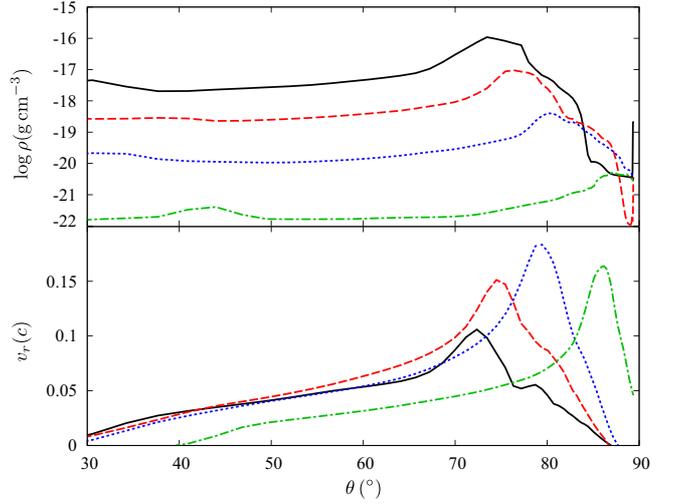}
  \end{center}
  \caption{Angular profiles of the density (top panel) 
and the radial velocity (bottom panel) at the outer boundary.
The black solid, red dashed, blue dotted, and green dashed-dotted lines show the profiles for
$M_{\rm BH}=10^6 M_{\odot}$,
$10^7 M_{\odot}$, $10^8 M_{\odot}$, and $10^9 M_{\odot}$. Here, the Eddington ratio is fixed to be $\varepsilon=0.1$. 
}\label{fig5}
\end{figure}

Such differences are explicitly shown in Fig. \ref{fig5},
in which we plot the angular profiles of the density (top panel) 
and the radial velocity (bottom panel) at the outer boundary
for $M_{\rm BH}=10^6 M_{\odot}$ (black solid lines), 
$10^7 M_{\odot}$ (red dashed lines), $10^8 M_{\odot}$ 
(blue dotted lines),
and $10^9 M_{\odot}$ (green dashed-dotted lines).
Here, the Eddington ratio is $\varepsilon=0.1$. 
This figure shows that 
the density and the radial velocity have peaks at 
$\theta \sim 72^\circ$ for $10^6 M_{\odot}$,
$\theta \sim 74^\circ$ for $10^7 M_{\odot}$,
$\theta \sim 78^\circ$ for $10^8 M_{\odot}$,
and $\theta \sim 86^\circ$ for $10^9 M_{\odot}$.

The top panel of Fig. \ref{fig5} shows that 
the density drastically decreases 
with the increase of the black hole mass. 
The peak density is 
$1.1\times 10^{-16}\,{\rm g\,cm^{-3}}$ for $10^6 M_{\odot}$,
$9.6\times 10^{-18}\,{\rm g\,cm^{-3}}$ for $10^7 M_{\odot}$,
$4.0\times 10^{-19}\,{\rm g\,cm^{-3}}$ for $10^8 M_{\odot}$,
and $5.0\times 10^{-21}\,{\rm g\,cm^{-3}}$ for $10^9 M_{\odot}$.
We find that the peak density is roughly 
proportional to $M_{\rm BH}^{-1}$.
In the bottom panel of Fig. \ref{fig5}, 
we find that the peak velocity slightly increases 
with an increase in the black hole mass as
$v_r \propto M_{\rm BH}^{1/8}$
in the range of $10^6 M_{\odot} \leq M_{\rm BH}\leq 10^8 M_{\odot}$.
However, the case of $M_{\rm BH}=10^9 M_{\odot}$ is an exception.
The peak velocity for $M_{\rm BH}=10^9 M_{\odot}$
is smaller than that for $M_{\rm BH}=10^8 M_{\odot}$.

The inverse proportional relation between
the density and the black hole mass 
is understood as follows:  
When the black hole mass changes
while maintaining the Eddington ratio constant,
the force multiplier in the launching region ($\sim 30$--$40R_{\rm S}$)
is kept almost constant.
This is because the gravity ($\propto M_{\rm BH}/r^2$)
and the radiation flux ($\propto L_{\rm D}/r^2$) 
have the same $M_{\rm BH}$-dependency ($\propto M_{\rm BH}^{-1}$).
At the launching region,
where the X-ray is effectively obscured ($\log \xi \ll 2$),
the force multiplier is approximately a function of 
the local optical depth parameter, $t \propto \rho |dv/ds|^{-1}$, 
although the force multiplier depends on 
$t$ and $\xi$ for $\log \xi \gtrsim 2$.
The distance $ds$ is proportional to $M_{\rm BH}$
and the velocity is roughly evaluated by the escape velocity
at the launching region ($dv\propto M_{\rm BH}^0$),
so that the relation of $\rho \propto M_{\rm BH}^{-1}$ is derived
from $t=$ constant.

\begin{figure}
  \begin{center}
    \includegraphics[width=\columnwidth]{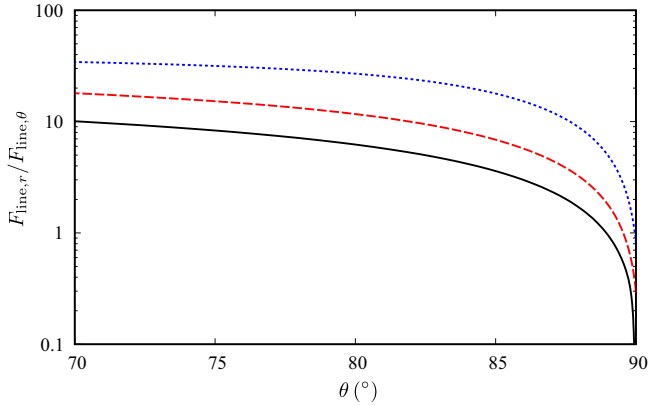}
  \end{center}
  \caption{
Ratio of the $r$-component to the $\theta$-component of the line-driving flux at $r=96R_{\rm S}$
as a function of the polar angle.
The black solid, red dashed, and blue dotted lines show the profiles for
$M_{\rm BH}=10^6 M_{\odot}$,
$10^7 M_{\odot}$, and $10^8 M_{\odot}$. 
Here, the Eddington ratio is fixed to be $\varepsilon=0.1$.
}\label{fig6}
\end{figure}

Next, we discuss the relation between the radial velocity 
and the black hole mass. 
Fig. \ref{fig6} shows the angular distribution of 
$F_{{\rm line},r}/F_{{\rm line},\theta}$
for $M_{\rm BH}=10^6 M_{\odot}$ (black solid line),
$10^7 M_{\odot}$ (red dashed line), and $10^8 M_{\odot}$ (blue dotted line)
at $r=96R_{\rm S}$, 
where the line-driving flux effectively contributes to the wind acceleration. 
We find that $F_{{\rm line},r}/F_{{\rm line},\theta}$ becomes large 
with the increase in the black hole mass. 
Since the effective temperature of the disk 
is proportional to $r^{-3/4} M_{\rm BH}^{-1/4}$,
the emission region of the line-driving radiation 
($200$--$3200\,{\rm \AA}$) 
tends to concentrate near the black hole,
as the black hole mass increases.
As a result, for the large black hole mass, 
the acceleration in the $r$-direction is more efficient 
than that in the $\theta$-direction 
and the radial velocity slightly increases 
as $v_r \propto M_{\rm BH}^{1/8}$.
Also, such an enhanced $F_{{\rm line},r}/F_{{\rm line},\theta}$ causes 
the wind with a large opening angle (see the bottom panel in Fig. \ref{fig5}).

In the top panel of Fig. \ref{fig5}, 
we find that the peak density for $M_{\rm BH}=10^9 M_{\odot}$
and $\varepsilon=0.1$ 
is smaller than expected from the 
relation of $\propto M_{\rm BH}^{-1}$.
This can be understood from 
the reduction in the effective temperature of the accretion disk.
The effective temperature is proportional to 
$\varepsilon M_{\rm BH}^{-1/4}$,
so that the line-driving luminosity 
decreases if the black hole mass is too large 
or the Eddington ratio is too small.
Thus, the line force 
for $M_{\rm BH}=10^9M_\odot$ and $\varepsilon=0.1$ 
is too small to produce the powerful disk winds.
As a result,
the density for $M_{\rm BH}=10^9M_\odot$
becomes one tenth or less
in comparison with the case of $M_{\rm BH}=10^8M_\odot$.
Here we note that, even $M_{\rm BH}=10^9M_\odot$,
such a reduction of the wind density is not remarkable
in the case of $\varepsilon=0.3$ or $0.5$,
since the disk temperature 
is not significantly low.
The bottom panel in Fig. \ref{fig5} indicates that,
in the case of $\varepsilon=0.1$,
the radial velocity is smaller 
for $M_{\rm BH}=10^9 M_{\odot}$
than for $M_{\rm BH}=10^8 M_{\odot}$.
This is also caused by the reduction in the line force
due to the decrease in the disk temperature.
Hence, the mass outflow rate, the momentum flux, and the
kinetic luminosity become small in the case of 
$M_{\rm BH}=10^9M_\odot$ and $\varepsilon=0.1$
(see Fig. \ref{fig3}).

From the resulting relations of 
$\rho \propto M_{\rm BH}^{-1}$ and 
$v_r \propto M_{\rm BH}^{1/8}$,
the $M_{\rm BH}$-dependence of 
$\dot{M}_{\rm w}$, $\dot{p}_{\rm w}$, and $\dot{E}_{\rm w}$
for given Eddington ratio roughly becomes
$d \log \dot{M}_{\rm w} / d\log {M}_{\rm BH} \sim 9/8$,
$d \log \dot{p}_{\rm w} / d\log {M}_{\rm BH} \sim 10/8$,
and $d\log \dot{E}_{\rm w}/ d\log {M}_{\rm BH} \sim 11/8$.
They are not inconsistent with the 
observed scaling of 
$d\log\dot{M_{\rm w}}/d\log{L_{\rm bol}}\sim 0.9$, 
$d\log\dot{p_{\rm w}}/d\log{L_{\rm bol}}\sim 1.2$, 
and $d\log\dot{E_{\rm w}}/d\log{L_{\rm bol}}\sim 1.5$.
This is why our model can explain the scaling relations of the UFOs.


\section{Discussion}
\label{discussions}
\subsection{Comparison with observations}

We successfully show that the line-driven disk wind explains 
the scaling relation of the mass outflow rate, 
momentum flux, 
and the kinetic luminosity with the AGN luminosity, 
which was discovered by G15.
However, in Fig. \ref{fig3}, 
the result for $\varepsilon=0.1$ and $M_{\rm BH}=10^9 M_{\odot}$
is located bellow the lower boundary of the blue shaded area.
This is induced by the reduction of the disk temperature
due to the large black hole mass and small Eddington ratio,
as we have mentioned above.
Such a discrepancy might disappear
if the wind is launched from the region of $r\lesssim 30 R_{\rm S}$,
since the line-driving radiation,
which is mainly emitted near the black hole,
enhances the wind power.
In the present simulations, 
we set the inner radius of the computational domain to be
$r_{\rm in}=30R_{\rm S}$
based on the assumption that the matter is fully ionized near the black
hole ($r\leq 30 R_{\rm S}$).
However, the size of the ionization region is still unknown.
If the ionized region shrinks for the case of 
$\varepsilon=0.1$ and $M_{\rm BH}=10^9 M_{\odot}$,
$\dot M_{\rm w}$, $\dot p_{\rm w}$, and $\dot E_{\rm w}$
would increase and overlap with the blue shaded area.

In our model, 
it is difficult for the line-driven mechanism to accelerate
the disk wind for small Eddington ratio,
implying that the UFOs are not observed in the AGNs 
with the small Eddington ratio.
This result is consistent with the observations of G15.
In our simulations, 
the line-driven wind does not appear for $\varepsilon\leq 0.01$.
In G15, the large sample of UFOs is detected
in AGNs with $\varepsilon\sim 0.1$,
and the UFOs are not detected in the AGN with $\varepsilon \lesssim 0.01$.
Although the Eddington ratio of NGC3227 is
estimated to be $\sim 0.003$,
this sample is excluded when the scaling relation
with the bolometric luminosity is discussed,
since the outflow velocity ($v_{\rm w}<0.008c$)
is much smaller than that of the typical UFOs ($v_{\rm w}\gtrsim 0.1c$)
and it is unclear whether the absorber is the part of the disk wind. 
Also, UFOs are detected in the four AGNs with 
$\varepsilon\sim L_{\rm bol}/L_{\rm Edd}\sim 0.025$, 
which is the smallest Eddington ratio in the UFO sample of G15,  
and their black hole masses are distributed
in $10^{7.1} M_{\odot}\lesssim M_{\rm BH}\lesssim 10^{7.6} M_{\odot}$. 
This is also consistent with our results,
whereby the wind is launched only for the case of
$M_{\rm BH}\lesssim 10^8 M_{\odot}$ when $\varepsilon=0.025$.

Although our model can roughly reproduce the scaling relation of G15,
the resulting $M_{\rm BH}-$dependence of the outflow velocity,
$v_{\rm w}\propto M_{\rm BH}^{1/8}$,
is different from that of G15, $v_{\rm w}\propto M_{\rm BH}^{1/2}$.
Such a discrepancy might be caused by the difference of
the measuring method.
The wind velocity of the UFOs is estimated from the blueshifted
absorption lines.
On the other hand, in our simulations,
the radial velocity of the main stream of the wind 
(peak velocity in the bottom panel of Fig. \ref{fig5})
is recognized as the wind velocity.
However, the absorption lines via the main stream
are not always observed.
If the high-velocity part of the wind is highly ionized,
the intrinsic outflow velocity of the wind cannot be observed.
In addition, the observed wind velocity would depend on
the observer's viewing angle.
As shown in Fig. \ref{fig5}, the wind velocity is sensitive to
the angle.
This problem should be solved by the calculation of the emergent spectra
(see below).

Finally,
the wide dispersion in the plot of $L_{\rm bol}$-dependence of
$\dot{M}_{\rm w}$, $\dot{p}_{\rm w}$, and $\dot{E}_{\rm w}$
(blue shaded area in Fig. \ref{fig3})
might originate from the Eddington ratio,
the observer's viewing angle,
and the ionization degree of the wind matter.
As we have already mentioned,
the lines for fixed black hole mass
(red dashed lines in Fig. \ref{fig3}) are
not in parallel with respect to the best-fitting lines.
Thus, the difference of the Eddington ratio
works to make a scatter in the plot.
In addition, the density as well as the velocity
is sensitive to the polar angle as we have mentioned above,
so that the estimated
$\dot{M}_{\rm w}$, $\dot{p}_{\rm w}$, and $\dot{E}_{\rm w}$
are expected to depend on the observer's viewing angle.
If the X-ray luminosity varies,
the ionization degree of the wind matter would vary.
If this is the case,
the observed absorption features
change, leading the scatter in the plot.
The calculation of the emergent spectra
should make the point clear.

\subsection{Radiation drag}
In our simulations,
the basic equations are non-relativistic and the radiation drag force, 
which is on the order of $v/c$, 
is not taken into consideration. 
However, the agreement between our model and the observations does not change by the radiation drag. 
In order to make sure of that, 
we perform a simulation in which the radiation force is artificially reduced by 40\%, 
since we find based on the numerical data of fiducial parameters 
that the maximum value of the radiation drag term is around 40\% of the radiation flux force 
(the radiation force shown by the equation \ref{radforce}) 
at the region of the velocity with $\sim 0.2c$. 
Here we note that such a treatment overestimates the effect of the radiation drag. 
This is because the radiation drag force would be very inefficient 
in the launching region where the velocity of the wind is small, $\ll c$. 

As a result of the case of the fiducial parameter set, 
we find that the velocity is not very different from the original value. 
The radial velocity of the main stream at the outer boundary is $\sim 0.18c$, 
which is almost the same as the original result 
(see the peak velocity in the bottom panel of Fig. \ref{fig5}).
The peak density at the outer boundary slightly deceases from the original value of 
$4\times 10^{-19}\, {\rm g\,cm^{-3}}$ to $7\times 10^{-20}\, {\rm g\,cm^{-3}}$. 
As a consequence, the mass outflow rate, momentum flux, and kinetic luminosity become 
$1.6\times 10^{24}\,{\rm g\,s^{-1}}$, $6.7\times 10^{33}\,{\rm g\, cm\, s^{-2}}$, 
and $1.5\times 10^{43}\,{\rm erg\, s^{-1}}$,
which are slightly smaller than the original value, 
$5.0\times 10^{24}\,{\rm g\,s^{-1}}$, $2.0\times 10^{34}\,{\rm g\, cm\, s^{-2}}$, 
and $4.7 \times 10^{43}\,{\rm erg\, s^{-1}}$. 
Here, the deviation would be smaller in reality, 
since the radiation drag is overestimated in this test calculation. 
Thus, we conclude that our results do not change so much even if the radiation drag is included in the simulations.

Although we simply investigate the effect of the radiation drag using the above method, 
the multi-color radiation hydrodynamics simulations would be necessary for the exact calculations. 
This is because the radiation drag via the line absorption would be sensitive to the spectral shape of the radiation
 as well as the Doppler shift.

\subsection{Future works}
The mass outflow rate estimated by our simulations is comparable to 
or slightly larger than the mass accretion rate for large Eddington ratio
($\varepsilon=0.3$ and 0.5).
The mass ejection from the accretion disk 
reduces the mass accretion onto the black hole,
and affects the disk luminosity as well as the
effective temperature of the disk.
Then, the wind structure might change.
However, the impact on the disk due to the launching of the wind
is not considered in our simulations.
We treat the emission and the density at the disk surface
as not changing with time.
Our method might overestimate the mass outflow rate of the wind.
In order to research the wind more realistically, 
it is important to perform the radiation hydrodynamics simulations
in which the disk and wind are self-consistently solved.
The numerical simulations
of the near-Eddington disks and outflows are performed by 
\citet{KO06,KO09,KO11},
but the line force is not included.
The simulations taking into account the line force
are remain future work.

In this paper, 
we calculate the mass accretion rate, momentum flux,
and kinetic luminosity, 
and show that their scaling relations with disk luminosity
are consistent with the results by G15.
However, for more precise verification of the model, 
it is necessary to investigate the absorption lines of 
FeX\hspace{-.1em}X\hspace{-.1em}V and/or 
FeX\hspace{-.1em}X\hspace{-.1em}V\hspace{-.1em}I
based on our simulation data.
The spectral synthesis based on the results of simulations is important
future work.
The synthetic spectra calculated based on the results of \citet{Proga04}
have been reported by \citet{Schurch09,Sim10,HP14}.

Recently, the time variation of the absorption lines
is detected
\citep[e.g.,][]{Misawa07,Cap13,Tombesi12b}. 
This implies that the wind structure changes in time and/or
the wind has a non-axisymmetrical structure. 
Although the density fluctuations of line-driven wind
in one- or two-dimensional calculations
have been reported by \citet{OP99,Proga00},
the non-axisymmetrical structure has not been investigated yet.
To reveal the origin of the time variability of the absorption lines,
we should perform time-dependent three-dimensional simulations.

\section{Conclusions}
\label{conclusions}
We performed two-dimensional radiation hydrodynamics simulations 
of line-driven disk winds in AGNs. 
The resulting scaling relations of 
the mass outflow rate ($\dot M_{\rm w}$),
momentum flux ($\dot p_{\rm w}$), 
and kinetic luminosity ($\dot E_{\rm w}$)
with the disk luminosity
are consistent with those obtained by X-ray observations (G15).

We found
$d\log\dot{M_{\rm w}}/d\log M_{\rm BH}\sim 9/8$, 
$d\log\dot{p_{\rm w}}/d\log M_{\rm BH}\sim 10/8$, 
and $d\log\dot{E_{\rm w}}/d\log M_{\rm BH}\sim 11/8$
for the range of $M_{\rm BH}=10^{6-9}M_\odot$ and $\varepsilon=0.1$--$0.5$
with $M_{\rm BH}$ and $\varepsilon$ being the
black hole mass and the Eddington ratio.
Since the Eddington ratio is limited within the narrow range of 
$\sim 0.1$--$0.5$ for launching the winds,
the above $M_{\rm BH}$-dependence can be replaced 
with the scaling relation with the bolometric luminosity
(
$d\log\dot{M_{\rm w}}/d\log{L_{\rm bol}}\sim 9/8$, 
$d\log\dot{p_{\rm w}}/d\log{L_{\rm bol}}\sim 10/8$, 
and $d\log\dot{E_{\rm w}}/d\log{L_{\rm bol}}\sim 11/8$
).
They are roughly consistent with 
the scaling relations obtained by the X-ray observations,
$d\log\dot{M_{\rm w}}/d\log{L_{\rm bol}}\sim 0.9$, 
$d\log\dot{p_{\rm w}}/d\log{L_{\rm bol}}\sim 1.2$, 
and $d\log\dot{E_{\rm w}}/d\log{L_{\rm bol}}\sim 1.5$.
At least, since there is large scatter in the UFO sample,
our results are in the region containing 
90\% of the posterior probability distribution 
in the $\dot M_{\rm w}$-$L_{\rm bol}$ plane, 
$\dot p_{\rm w}$-$L_{\rm bol}$ plane,
and $\dot E_{\rm w}$-$L_{\rm bol}$ plane (G15).

Our simulations also revealed that
the line-driven winds do not appear for the AGNs with
small Eddington ratio, $\varepsilon \lesssim 0.01$.
In addition, in the case of $\varepsilon = 0.025$,
the black hole mass is limited to be $\lesssim 10^8 M_{\odot}$
for launching the winds.
These results are also consistent with G15.
Indeed, in G15, the smallest Eddington ratio in the UFO sample
is $\sim 0.025$, and their black hole masses are
$10^{7.1} M_{\odot}\lesssim M_{\rm BH}\lesssim 10^{7.6} M_{\odot}$.
There are no AGNs with $\varepsilon \leq 0.01$ exhibiting UFO features
in the sample of G15.
In their sample, although NGC3227 has very small Eddington ratio
($\sim 0.003$), the outflow velocity of this object
is very small, $<0.008c$, which is far from the UFOs.

The number of UFO samples is still small, 
but the line-driven disk winds successfully explain 
the scaling relations of the mass outflow rate, momentum flux,
and the kinetic luminosity with AGN luminosity. 
We conclude that the line-driven disk wind
is a plausible model for UFOs in AGNs.

\section*{Acknowledgements}

We would like to thank Hiroyuki R. Takahashi for useful discussions. Numerical computations were carried out on Cray XC30 at the Center for Computational Astrophysics, National Astronomical Observatory of Japan. 
This work is supported in part by JSPS Grant-in-Aid for Scientific Research (C) (15K05036 K.O., 16K05309 K.E.). This work was supported in part by MEXT and JICFuS as a priority issue (Elucidation of the fundamental laws and evolution of the universe) to be tackled by using Post K Computer.








\bsp	
\label{lastpage}
\end{document}